\documentclass[3p,times,twocolumn]{elsarticle}

\usepackage{ecrc}

\volume{00}

\firstpage{1}

\journalname{Nuclear and Particle Physics Proceedings}

\runauth{}

\jid{nppp}

\jnltitlelogo{Nuclear and Particle Physics Proceedings}

\usepackage{amssymb}

\usepackage[figuresright]{rotating}

\begin{document}

\begin{frontmatter}



\dochead{}

\title{Electromagnetic radiation and collectivity in small quark-gluon droplets}


\author[1]{Chun Shen}
\author[3]{Jean-Fran\c{c}ois Paquet}
\author[4]{Gabriel S. Denicol}
\author[2]{Sangyong Jeon}
\author[2]{Charles Gale}
\address[1]{Physics Department, Brookhaven National Laboratory, Upton, NY 11973, USA}
\address[2]{Department of Physics, McGill University, 3600 University Street, Montreal, QC, H3A 2T8, Canada}
\address[3]{Department of Physics and Astronomy, Stony Brook University, Stony Brook, New York 11794, USA}
\address[4]{Instituto de F\'{i}sica, Universidade Federal Fluminense, UFF, Niter\'{o}i, 24210-346, RJ, Brazil}

\begin{abstract}
We study the multiplicity and rapidity dependence of thermal and prompt photon production in p+Pb collisions at 5.02 TeV, using a (3+1)D viscous hydrodynamic framework. Direct photon anisotropic flow coefficients $v^\gamma_{2,3}$ and nuclear modification factor $R^\gamma_\mathrm{pPb}(p_T)$ are presented in both the p-going (backward) and the Pb-going (forward) directions. The interplay between initial state cold nuclear effect and final state thermal enhancement at different rapidity regions is discussed. 
The proposed rapidity dependent thermal photon enhancement and direct photon anisotropic flow observables can elucidate non-trivial longitudinal dynamics of hot quark-gluon plasma droplets created in small collision systems.
\end{abstract}


\end{frontmatter}


\section{Introduction}
\label{intro}

Recent measurements of high energy proton-nucleus collisions at the Large Hadron Collider (LHC) show intriguing evidence of collective behaviour which suggest the creation of hot and strongly coupled quark-gluon plasma (QGP) \cite{Chatrchyan:2013nka,Aad:2013fja,ABELEV:2013wsa}. Such collisions offer the opportunity to study QCD (Quantum Chromodynamics) under extreme conditions. Theoretical analyses using viscous hydrodynamics can provide a reasonable description of the measured hadronic flow observables \cite{Bozek:2013uha,Nagle:2013lja,Shen:2016zpp}. Understanding the origin of such a strong collectivity in small collision systems is currently a very active topic in the field. In order to quantify the properties of the QCD matter produced in these collisions, one needs to study multiple aspects of experimental observables within a consistent framework.Thus, in addition to hadronic flow observables, we will study penetrating electromagnetic probes from these small collision systems.

Electromagnetic observables, such as direct photons, are regarded as clean probes to the produced matter in heavy-ion collisions \cite{Shen:2015nto,Shen:2016odt}. Owing to the small system size in p+Pb collisions, the produced photons can escape the medium without any further interaction. Hence, they carry undistorted information from their local production points to the detectors. 

This work studies the multiplicity and rapidity dependence of direct photon observables in p+Pb collisions at 5.02 TeV using event-by-event (3+1)D viscous hydrodynamic simulations. The results presented here provide  information complementary to that in our systematic study of Refs. \cite{Shen:2016zpp,Shen:2015qba}.

\section{Results and Discussion}
\label{results}
\begin{figure*}[t!]
\centering
\begin{tabular}{cc}
\includegraphics[width=0.45\textwidth]{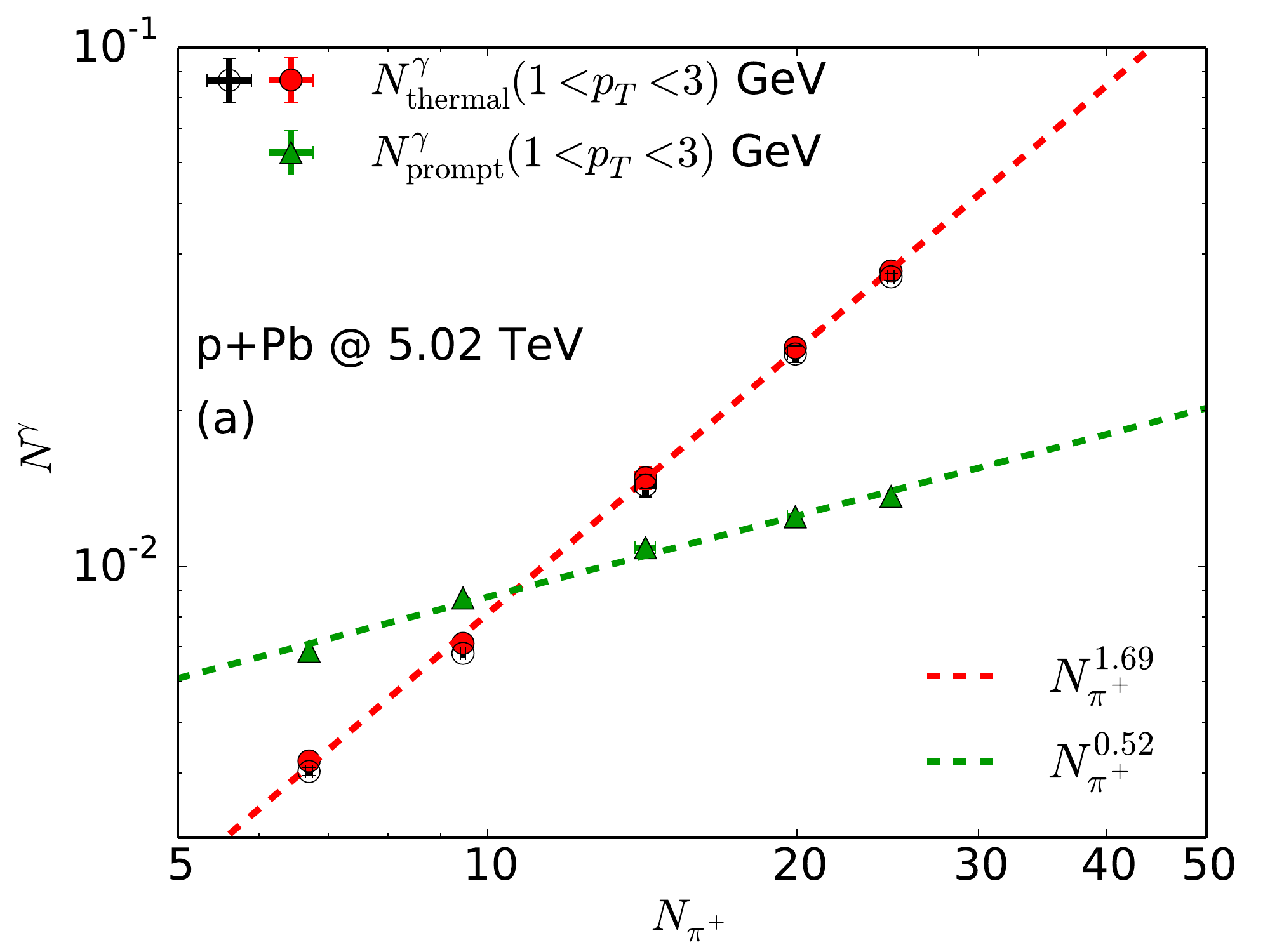}
\includegraphics[width=0.45\textwidth]{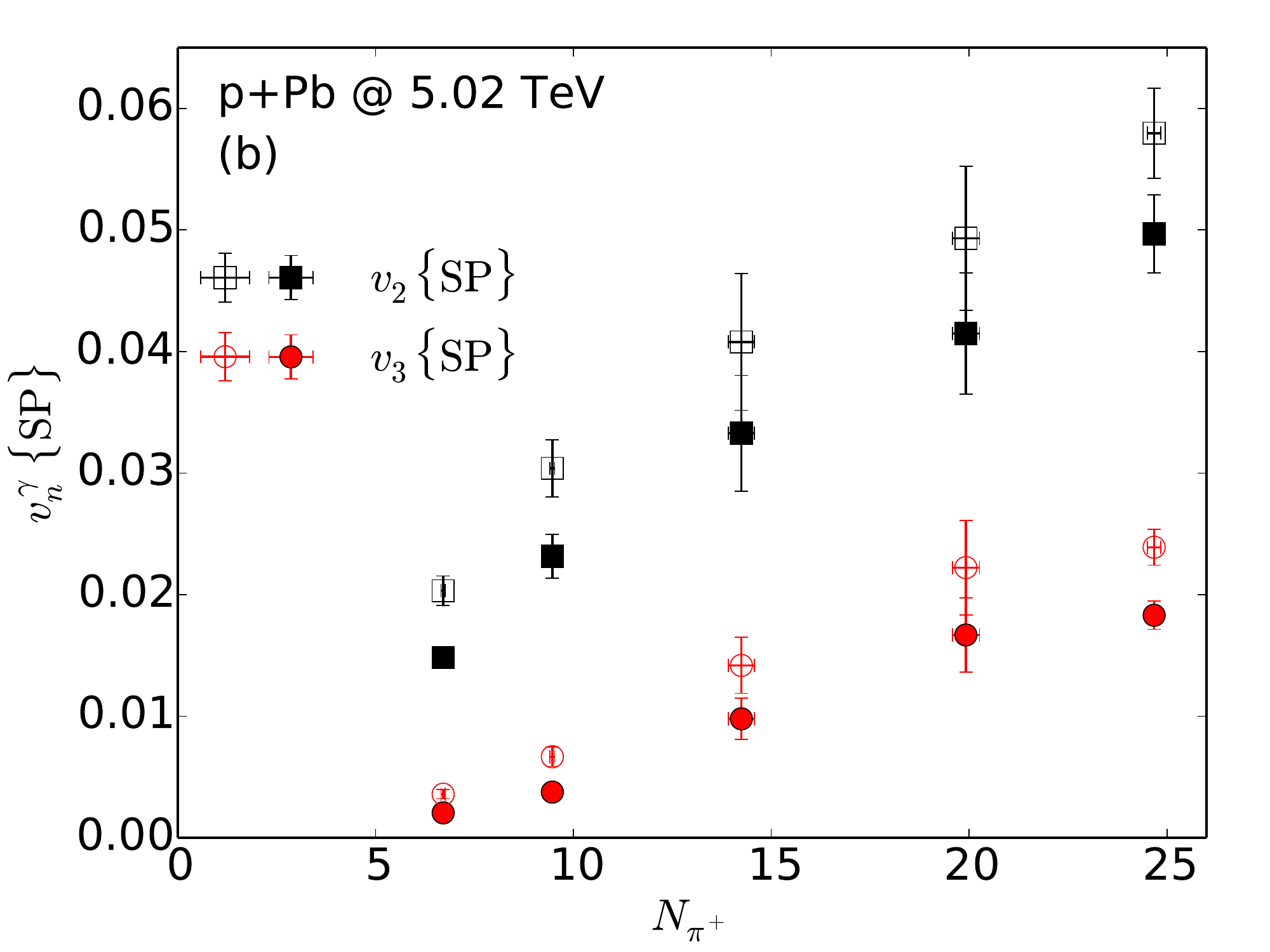}
\end{tabular}
\caption{{\it Panel (a):} Multiplicity dependence of thermal and prompt photon yields with transverse momentum between 1 and 3 GeV in p+Pb collisions at 5.02 TeV. {\it Panel (b):} Direct photon anisotropic flow coefficients integrated from 1 to 3 GeV, $v_{2,3}\{\mathrm{SP}\}$, as a function of pion multiplicity in p+Pb collisions at 5.02 TeV. Open markers show results without $\delta f$ correction to photon emission rates.}
\label{fig1}
\end{figure*}

The event-by-event dynamical evolution of p+Pb collision at 5.02 TeV is modelled using a hydrodynamics + hadronic cascade hybrid framework \cite{Shen:2016zpp}. The Monte-Carlo Glauber model is employed to generate fluctuating initial conditions. Collision-by-collision multiplicity fluctuations are included to reproduce the measured charged hadron multiplicity distributions \cite{Shen:2014vra}. The rapidity dependence of the initial energy density profile is parametrized with envelope functions \cite{Bozek:2011if,Shen:2016zpp}. The individual energy density profile is then evolved using (3+1)D viscous hydrodynamics MUSIC \cite{Schenke:2010nt} coupled with a lattice based equation of state s95p-v1.2 \cite{Huovinen:2009yb}. Individual fluid cells are converted to particles at an isothermal hyper-surface, where the switching temperature is  $T_\mathrm{sw} = 155$ MeV. Particles are fed into a hadronic transport model (UrQMD) for further scatterings and decays. We choose a specific shear viscosity of the QGP phase $\eta/s = 0.1$, to reproduce the hadronic flow observables (see Ref. \cite{Shen:2016zpp}).
For direct photon production, we consider contributions from prompt and thermal radiation. Pre-equilibrium photon contributions from a saturated gluon phase \cite{Benic:2016uku} are not considered here. The prompt photons are evaluated with perturbative QCD at next-to-leading order in $\alpha_s$ \cite{Shen:2016zpp}. The isospin effect is included for the Pb nucleus. Cold nuclear effects are taken into account with the nCTEQ15 nuclear parton distribution functions \cite{Kovarik:2015cma}. Thermal radiation is computed by folding thermal photon emission rates with the evolving hot medium. The details of the calculations can be found in Ref.~\cite{Shen:2016zpp}.

\begin{figure*}[ht!]
\centering
\begin{tabular}{cc}
\includegraphics[width=0.45\textwidth]{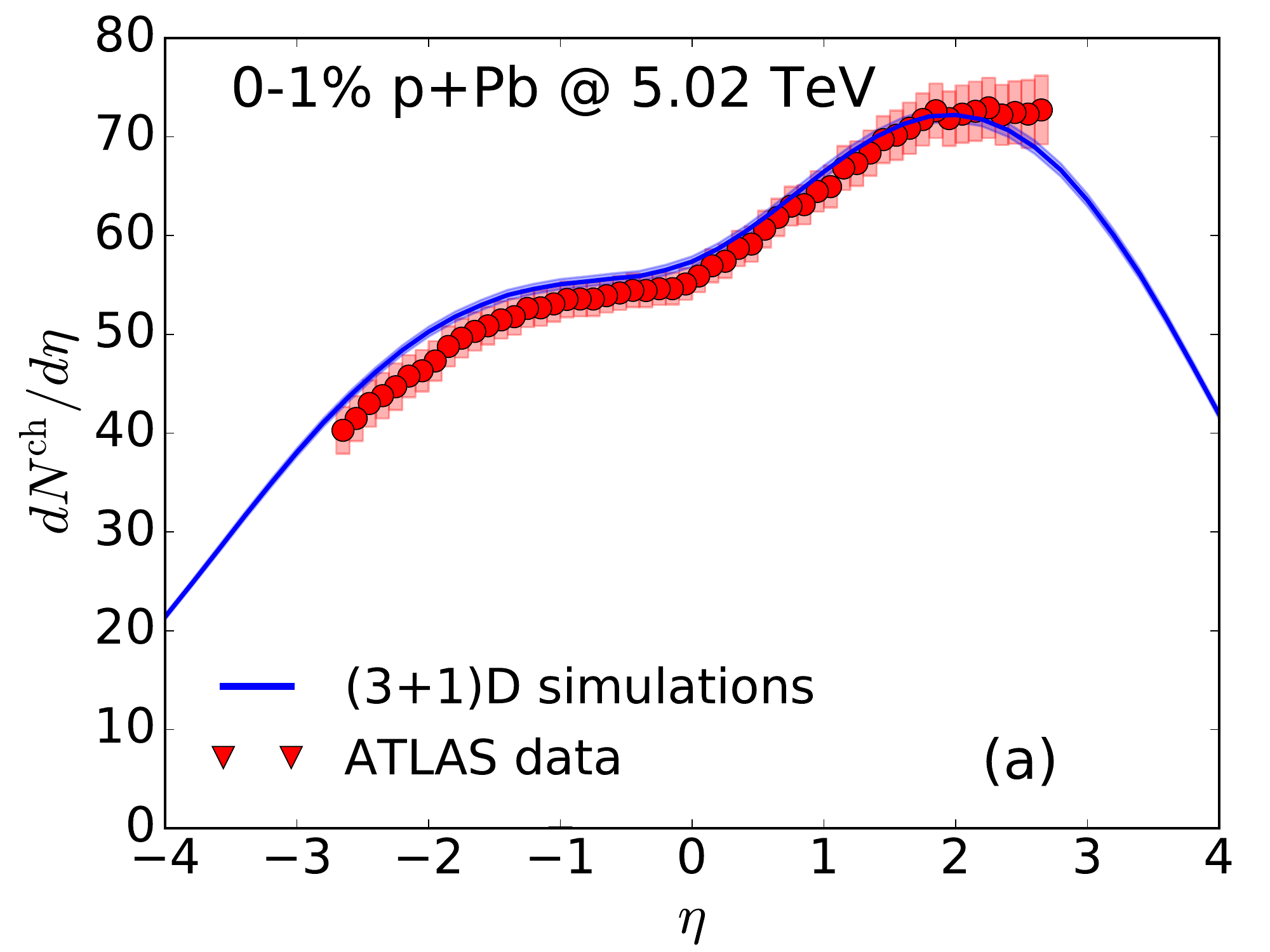}
\includegraphics[width=0.45\textwidth]{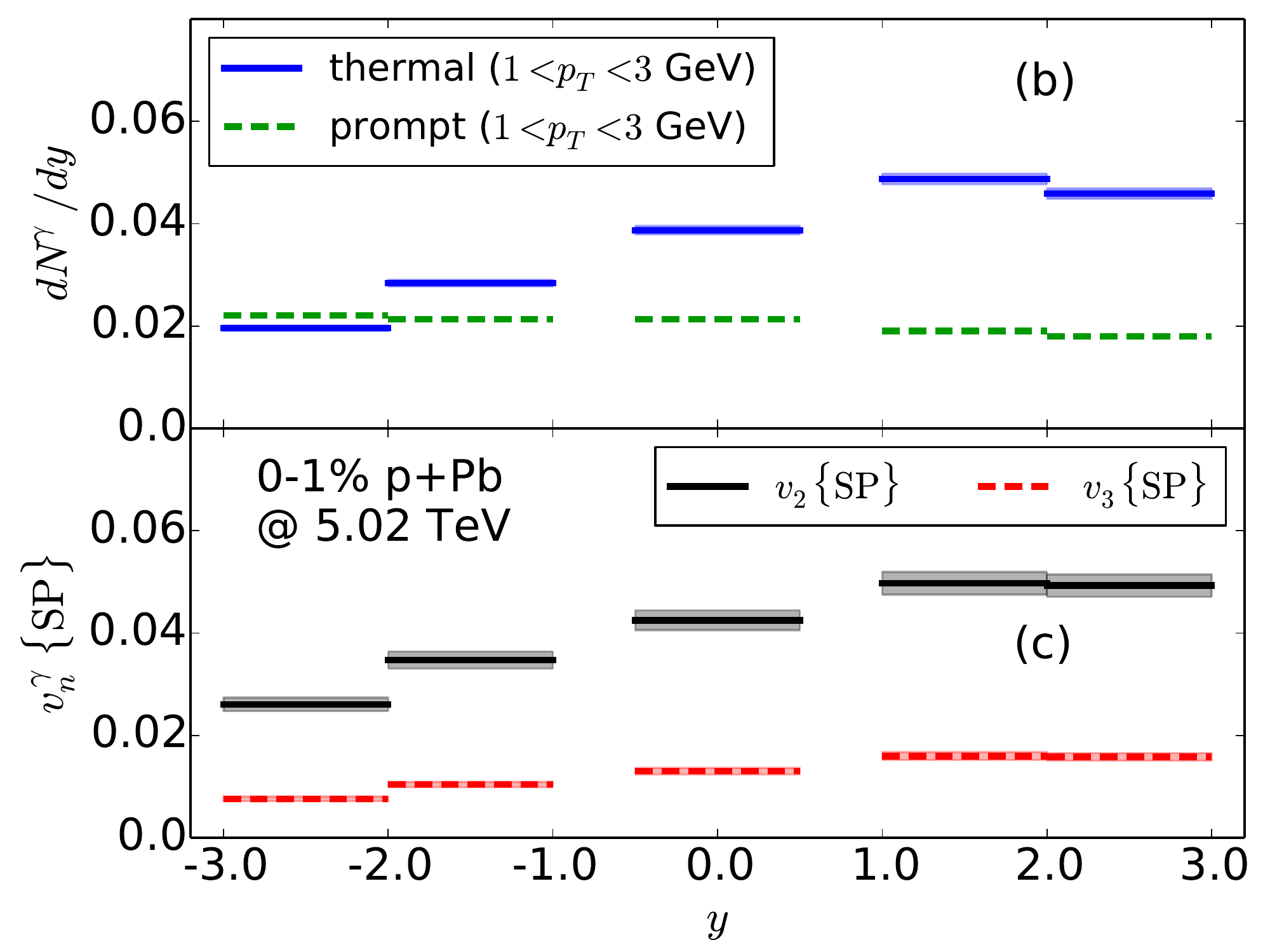}
\end{tabular}
\caption{{\it Panel (a):} Charged hadron multiplicity as a function of pseudorapdity compared with the ATLAS measurements \cite{Aad:2015zza} in 0-1\% p+Pb collisions at 5.02 TeV. Positive rapidity is the Pb-going direction. {\it Panel (b):} The rapidity dependence of thermal and prompt photon yields in 0-1\% p+Pb collisions at 5.02 TeV. {\it Panel (c):} The $p_T$-integrated direct photon $v_n$ as a function of photon rapidity. In Panels (b) and (c), the transverse momentum is integrated from 1 to 3 GeV. }
\label{fig2}
\end{figure*}

We perform numerical simulations of p+Pb collisions from central to 50\% in collision centrality.
Fig.~\ref{fig1}a shows the prompt and thermal photon yields as a function of the final pion multiplicity. Good power law scaling with $N_\pi$ is found for the individual prompt and thermal component. The prompt photon yield increases slower than $N_\pi$ in p+Pb collisions, with a power exponent $\sim0.5$. It is smaller than its value in large symmetric nucleus-nucleus collisions, where $N^\gamma_\mathrm{prompt} \propto N_\pi^{1.2}$ \cite{Paquet:2016dry}.
This slow growth of prompt photon yield helps the thermal photon radiation to shine out in the total direct photon signal in central p+Pb collisions. The thermal photon yield goes like $N_\pi^{1.69}$, much faster than the prompt source. This can be understood as a superposition of the effects from an  increase of the system's average temperature and of space-time volume as one goes to central collisions \cite{Shen:2016zpp}. In Fig.~\ref{fig1}a, the difference between the open black circles and red filled points shows that the out-of-equilibrium correction to the photon rates coming from shear viscosity, usually denoted as $\delta f$, does not affect the $p_T$-integrated photon yields.
The multiplicity dependence of direct photon anisotropic flow coefficients $v^\gamma_{2,3}$ is presented in Fig.~\ref{fig1}b. Compared to charged hadron $v^\mathrm{ch}_n$, direct photon $v^\gamma_{2,3}$ shows a stronger multiplicity dependence. In the low multiplicity events $N_\pi < 10$, prompt photons dominate the total direct photon signal. Since they don't carry any momentum anisotropy in our framework, the direct photon $v^\gamma_n$ is strongly suppressed. As one goes to central p+Pb collisions, the thermal radiation wins over the prompt contribution in the direct photon signal. The absolute values of direct photon $v^\gamma_{2,3}$ are comparable to the charged hadron $v^\mathrm{ch}_n$.
Finally, by comparing the open symbols with filled points, we estimate that the $\delta f$ corrections to the photon emission rates from shear viscosity suppress the direct photon $v^\gamma_{2,3}$ by about 10-20\%.

Because of the asymmetry of the colliding nuclei, p+Pb collisions break the longitudinal boost-invariance explicitly. They provide an opportunity to study non-trivial longitudinal dynamics with direct photons. Fig.~\ref{fig2}a shows the pseudo-rapidity distribution of the charged hadron multiplicity $dN^\mathrm{ch}/d\eta$ in 0-1\% p+Pb collisions. The longitudinal profile for the initial energy density in the MC-Glauber initial condition is adjusted to reproduce ATLAS measurements \cite{Aad:2015zza}. Since there are more participant nucleons from the Pb nucleus, more energy is deposited to the forward rapidity region compared to the backward p-going direction. Thus, the fireball in the forward direction achieves a higher peak temperature at the initial time and lives longer compared the one in the backward rapidity region. This longitudinal asymmetry of energy deposition directly imprints itself in the rapidity dependence of the thermal photon production $dN^\gamma_\mathrm{thermal}/dy$, as shown in Fig.~\ref{fig2}b. The thermal photon $dN^\gamma_\mathrm{thermal}/dy$ has a similar shape as that of the charged hadron $dN^\mathrm{ch}/d\eta$.
The prompt photon yield is slightly suppressed in the Pb-going direction compared to its value at mid-rapidity. This is caused by the nuclear shadowing effect from the Pb nuclear PDF.
This shadowing effect is reduced in the backward p-going direction. So with such an interplay between thermal and prompt photons in the different rapidity regions, the thermal photon radiation shines over the prompt photon background in the Pb-going direction. The maximum thermal to prompt ratio can be reached in the rapidity region $1 < y < 2$.

Fig.~\ref{fig2}c further highlights the $p_T$-integrated direct photon anisotropic flow coefficients as a function of rapidity. The direct photon $v^\gamma_{2,3}(y)$ also reaches a maximum in the rapidity region $1 < y < 2$ in the Pb-going direction where the thermal photon production peaks. Because of the dilution by prompt photons, the direct photon $v^\gamma_{2,3}$ has a stronger rapidity dependence than that of charged hadron $v^\mathrm{ch}_n(\eta)$ shown in Ref.~\cite{Shen:2016zpp}. 

\begin{figure}[h!]
\centering
\includegraphics[width=0.45\textwidth]{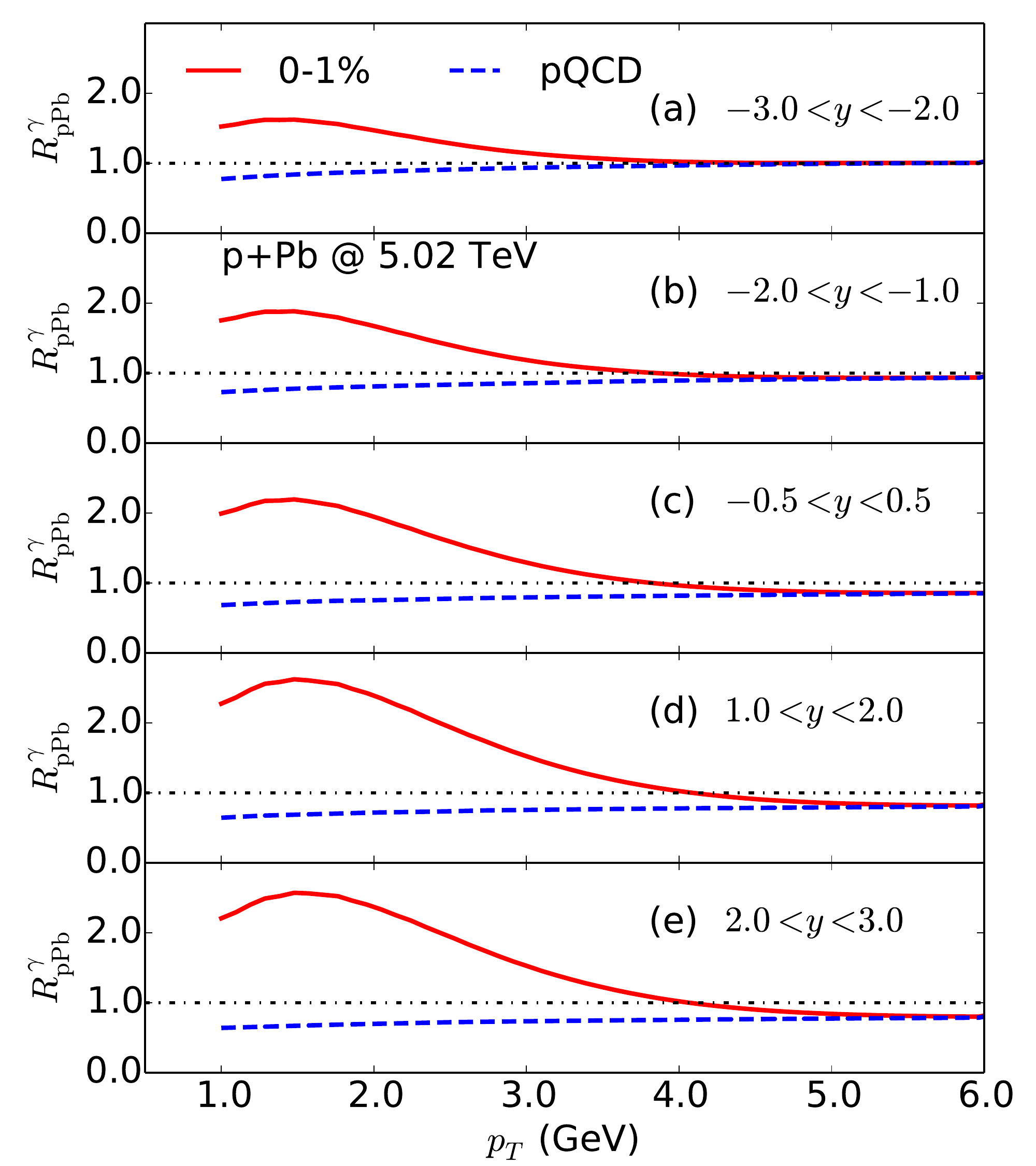}
\caption{The direct photon nuclear modification factor $R^\gamma_\mathrm{pPb}(p_T)$ in 0-1\% p+Pb collisions at three different rapidity regions. Positive rapidity denotes the Pb-going side.}
\label{fig3}
\end{figure}

Finally, the direct photon nuclear modification factor $R^\gamma_\mathrm{pPb}$ in different rapidity regions is summarized in Fig.~\ref{fig3}. In central 0-1\% p+Pb collisions, a thermal enhancement in $R^\gamma_\mathrm{pPb}$ can be observed in the region where $1 < p_T < 3$ GeV. The enhancement is larger in the forward rapidity region owing to the hotter and longer lived fireball. The maximum $R^\gamma_\mathrm{pPb}$ can reach up to $\sim2.5$ in the rapidity window $1 < y < 2$.

\section{Conclusion}
\label{conclusion}

We study the multiplicity and rapidity dependence of thermal photon radiation in p+Pb collisions at 5.02 TeV. Both direct photon yield and its anisotropic flow coefficients shows a strong multiplicity dependence. In central p+Pb collisions, thermal photons can leave a measurable signal in direct photon observables with $1 < p_T < 3$ GeV. 

We found an  interesting and opposite rapidity dependence for thermal and prompt photon production in p+Pb collisions. Thermal enhancement in the total direct photon signal is expected to be more pronounced in the forward Pb-going region compared to the mid-rapidity and p-going direction. 
The medium created in $1 < y < 2$ is hotter, and lives for a longer lifetime when compared with other rapidity windows. Prompt photon production in this rapidity range is slightly suppressed by the larger nuclear shadowing effect in the Pb nucleus. Future measurements of the rapidity dependence of direct photon observables can therefore help us to further understand the longitudinal dynamics of p+Pb collisions.




\section*{Acknowledgement}
This work was supported in part by the U.S. Department of Energy, Office of Science under contract No. DE- SC0012704 and the Natural Sciences and Engineering Research Council of Canada. C.S. gratefully acknowledges a Goldhaber Distinguished Fellowship from Brookhaven Science Associates, and C. G. gratefully acknowledges support from the Canada Council for the Arts  through its Killam Research Fellowship program. J.-F.P. was supported by the U.S. D.O.E. Office of Science, under Award No. DE-FG02-88ER40388.  
Computations were made in part on the supercomputer Guillimin from McGill University, managed by Calcul Qu\'ebec and Compute Canada. The operation of this supercomputer is funded by the Canada Foundation for Innovation (CFI), NanoQu\'ebec, RMGA and the Fonds de recherche du Qu\'ebec - Nature et technologies (FRQ-NT). 
\bibliographystyle{elsarticle-num}
\bibliography{Shen_C}







\end{document}